\documentclass[aps,nofootinbib,floatfix,superscriptaddress]{revtex4} 
\usepackage{graphicx,bm}
\input epsf
\newcommand{\sfig}[2]{
\includegraphics[width=#2]{#1}
        }

\newcommand*{\rf}[1]{Fig. \ref{fig:#1}}
\newcommand*{\re}[1]{Eq. \ref{eq:#1}}

\def\cmm2{{\,\rm cm^{-2}}}
\def\cm2{{\,{\rm cm}^2}}
\def\cmm3{{\,{\rm cm}^{-3}}}
\def\gcmm3{{\,{\rm g\,cm^{-3}}}}

\def\etal{{\it et al. }}

\def\fun#1#2{\lower3.6pt\vbox{\baselineskip0pt\lineskip.9pt
  \ialign{$\mathsurround=0pt#1\hfil##\hfil$\crcr#2\crcr\sim\crcr}}}

\def\be{\begin{equation}}
\def\ee{\end{equation}}
\def\bea{\begin{eqnarray}}
\def\eea{\end{eqnarray}}

\newcommand*{\chisq}{\chi^{2}}
\newcommand*{\lcdm}{$\Lambda$CDM}
\newcommand*{\LCDM}{$\Lambda$CDM }

\begin{document}
\title{What Do We Really Know About Cosmic Acceleration?}

\author{Charles A. Shapiro}
\affiliation{Department of Physics, The
University of Chicago, Chicago, IL~~60637-1433}
\affiliation{Kavli Institute for Cosmological Physics, The
University of Chicago, Chicago, IL~~60637-1433}

\author{Michael S. Turner}
\affiliation{Kavli Institute for Cosmological Physics, The
University of Chicago, Chicago, IL~~60637-1433}
\affiliation{Department of Astronomy \& Astrophysics, The
University of Chicago, Chicago, IL~~60637-1433}

\date{\today}

\begin{abstract}
Essentially all of our knowledge of the acceleration history of the Universe -- including the acceleration itself -- is predicated upon the validity of general relativity.  Without recourse to this assumption, we use SNeIa to analyze the expansion history and find (i) very strong ($5\sigma$) evidence for a period of acceleration, (ii) strong evidence that the acceleration has not been constant, (iii) evidence for an earlier period of deceleration and 
(iv) only weak evidence that the Universe has not been decelerating since $z \sim 0.3$.
\end{abstract}

\maketitle

\section{Introduction}

A still puzzling feature of our Universe is its accelerated expansion.  Measurements of type Ia supernovae (SNeIa) provide direct evidence \cite{Riess:1998cb,Perlmutter:1998np,Astier:2005qq} and CMB anisotropy measurements provide important indirect evidence \cite{Tegmark:2003ud}.  Within the context of general relativity, the acceleration can be explained if the present mix of matter and energy in the Universe includes a dark energy component that contributes about 2/3 of the critical density.  Dark energy is defined by its large negative pressure, $p\sim-\rho$, and nearly spatially uniform distribution \cite{Turner:1998ex}.

While the energy of the quantum vacuum, mathematically
equivalent to a cosmological constant, is the simplest explanation for the 
dark energy and cause of cosmic acceleration, all attempts
to compute its numerical value have been unsuccessful, either leading to
divergent results or numbers that are orders of magnitude too large; this is
known as the cosmological constant problem \cite{Carroll:2000fy}.  Because it is possible that the energy of the quantum vacuum is zero or too small to explain cosmic acceleration, theoretical
physicists have explored a plethora of possibilities for the
dark energy, from a very light scalar field \cite{Ratra:1987rm,Wetterich:1987fm,Frieman:1995pm,Coble:1996te,Caldwell:1997ii} to the influence of extra
dimensions \cite{Deffayet:2001pu}, and have even considered the possibility that there is no need for dark energy and that it arises due to heretofore neglected ``ordinary'' physics within general relativity \cite{Kolb:2005da}.  A fair summary of the present state of affairs is that there is little understanding of
why the Universe is accelerating.

Given that state of affairs, it is not unreasonable to think more broadly, even entertaining the possibility that the
explanation may well have to do with gravity theory itself
\cite{Carroll:2003wy,Capozziello:2003tk,Freese:2002sq,Arkani-Hamed:2002fu,Dvali:2003rk}.  After all, we are confident that Einstein's theory must be modified to make it compatible with quantum mechanics, and perhaps cosmic acceleration has something to do with that modification.  However, relaxing the assumption of general relativity means
that we must abandon the Friedmann equation, which relates the expansion
rate to the matter and energy densities.  We might then wonder what
we really know about cosmic acceleration -- or if the Universe is {\it really} accelerating since almost everything we have inferred about it depends upon the assumption of the existence of dark energy.

To be specific, dark energy is often described as a smooth energy component with constant equation-of-state ($w=p/\rho\,=$\,const\,$\sim -1$), where quantum vacuum energy is the special case $w = -1$.  The acceleration history of the Universe is then simply fixed by the matter content ($\Omega_M$) and the value of $w$. It begins with a period of decelerated expansion and then crosses over to accelerated expansion at redshift
$$z_c = \left[\frac{\Omega_M}{(3w+1)(\Omega_M -1)}\right]^{1/3w}-1\sim 0.5\,;$$
see \rf{allqplot}.  Both $w$CDM with $w =-1.05 \pm 0.14$ and $\Omega_M =0.29\pm0.03$ and $\Lambda$CDM with
$\Omega_M =0.30 \pm0.05$ are reasonable fits to the SNe data and a wealth of other cosmological data \cite{Tegmark:2003ud}.  
(This in fact is the basic evidence for a period of accelerated expansion.)  The existence of an early period of decelerated expansion is also supported by the success of the
gravitational instability theory of structure formation and of big-bang
nucleosynthesis.  In particular, unless the equations governing the growth
of small density perturbations are radically different, it is not possible
to grow structure from small perturbations in an accelerating universe \cite{Turner:1998ex}.  
In addition, it is difficult for a universe that accelerates during BBN to reproduce the observed light element abundances unless one artificially adjusts with the cosmic neutrino asymmetry \cite{Carroll:2001bv}.

In this paper we relax the assumption of general relativity, but retain the weaker assumption of an isotropic and homogeneous space-time (which is supported by ample observational data) described by a metric theory of gravity (because there are few if any viable non-metric theory alternatives).  In this case, SNeIa still provide a direct kinematic probe of the expansion, and we show that we can directly infer the following:  (i) very strong ($5\sigma$) evidence for a period of accelerated expansion; (ii) strong evidence that the acceleration has not been constant; (iii) evidence for an earlier period of deceleration; 
and (iv) only weak evidence that the Universe has not been decelerating since $z\sim 0.3$.

\begin{figure}[t]
  \sfig{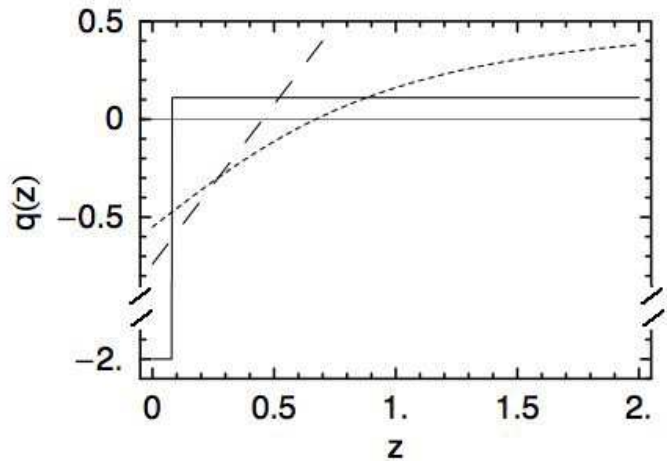}{.5\columnwidth}
  \caption{{\small Different parametrizations of $q(z)$.  The solid line is the best-fit 2-epoch model (\re{q2epoch}); the long-dashed line is the best-fitting linear model; the short-dashed line is the best-fitting \LCDM model ($\Omega_M=0.31$).  The kinematic models suggest a later transition to acceleration than \LCDM does. }}
  \label{fig:allqplot}
\end{figure}

\section{Friedmannless Cosmology} \label{sec:prelim}
\subsection{Kinematics}

With only the assumption that space-time can be described by a metric theory that is on the large isotropic and
homogeneous, the Robertson-Walker metric still pertains,
\be
ds^2 = -dt^2 + R(t)^2\left[ dr^2 + r^2 d\Omega \right] \,,
\ee
where $R(t)$ is the usual cosmic scale factor and for simplicity spatial
flatness has been assumed.  Likewise, the cosmic scale factor at a given epoch is related in the usual way to the redshift that a free-streaming photon has
suffered since that epoch: $R(t) = 1/(1+z)$.  The epoch dependent expansion and deceleration rates are defined as usual and are related to one another by
\begin{eqnarray}
H(z) & \equiv & \frac{\dot R}{R} \\
q(z) & \equiv & -\frac{\ddot R}{R H^2} = \frac{d}{dt}H^{-1} - 1 \\
H(z) & = & H_0 \exp \left[ \int_0^z [1+q(u)] d\ln (1+u) \right].
\end{eqnarray}

The relationship between detected energy flux ${\cal F}$ of a source at redshift $z$ and its intrinsic luminosity ${\cal L}$ is also unaffected, 
\bea
{\cal F} & \equiv & {{\cal L} \over 4\pi d_L^2} \\
d_L & = & (1+z) \int_0^z\,\frac{du}{H(u)} \; ,
\eea
where $d_L$ is the luminosity distance.  Defining the distance modulus $\mu$ for a standard candle (or standardizeable candle, e.g., SNeIa) in the usual way, 
\be
\mu (z) \equiv [m_B(z) - M_B] = 5\log (d_L/{\rm Mpc})  + 25 \label{eq:mu}
\ee
the observable, $\mu (z)$, is related to the acceleration history, $q(z)$,
\be
\mu (z) = 25 + 5\log \left [\frac{1+z}{H_0\cdot{\rm Mpc}} \int_0^z\,du \, \exp
\left(-\int_0^u\,[1+q(v)]d\ln v \right)  \right] \label{eq:mufromq}
\ee
Here $M_B$ and $m_B(z)$ are respectively the absolute and apparent
magnitudes of the source.  \re{mufromq} provides the fundamental kinematic relation that relates the deceleration history to SNeIa measurements. 

\subsection{Simple Kinematic Models}\label{sec:models}

Even with the structure provided by the Friedmann equation, it has been a struggle to find a generic way to describe dark energy; most have settled upon using the equation-of-state $w$.  Relaxing the assumption of the Friedmann equation exacerbates the problem.  The kinematic parameters, $H$ and/or $q(z)$ have been used; and it has been suggested \cite{Blandford:2004ah} that {\it jerk}, $j= -(\dddot a /a)/H^3$, might be a good kinematic choice since the case $j=1$ corresponds to a model that goes from matter-dominated behaviour to consmological constant dominated behaviour (of course, there is good evidence that our universe began from a radiation dominated phase which might have even been proceeded by a de Sitter phase).  Operationally, $H$, $q$ and $j$ are all equivalent since each one can be expressed using integrals or derivatives of the others.

Since $q$ is related to the equation-of-state in general relativity, $q = [\Omega_M + 3w(1-\Omega_M)]/2$, and there doesn't seem to be strong motivation to go beyond $q$, we focus here on the acceleration history, $q(z)$. We will consider three approaches to parameterizing $q(z)$:  (i) linear expansion, $q(z) = q_0 + zdq/dz$; (ii) piecewise constant; and (iii) a principal component analysis.  For all these analyses we will use the gold set of SNeIa culled by Reiss \etal which contains 157 well studied SNeIa between $z=0.1$ and $z=1.76$ \cite{Riess:2004nr}. 

Reiss \etal analyzed their own gold set using the linear expansion and found that $q_0<0$ at the 99.2\% level, that $dq/dz>0$ at the 99.8\% level and that the error ellipse for these two parameters is highly degenerate.
Based upon this they concluded that a transition from recent acceleration to past deceleration occurred at a redshift of $z_t=0.46\pm 0.13$.  Here we add that the best-fit parameters give a 12\% goodness-of-fit ($q_0=-.74$, $dq/dz=1.6$, $\chisq/{\rm DOF}=175/154$), on par with the best \LCDM model, which has an 11\% goodness-of-fit ($\Omega_M=0.31$, $\chisq/{\rm DOF}=177/155$).  The fact that this model finds $q_0<0$ gives us confidence that the Universe has accelerated recently.  Furthermore, the positive slope is certainly an indication that $q(z)$ was higher in the past.  But can we trust that there was ever a transition?  Unfortunately, this linear model {\it always} implies a transition redshift when its two parameters have opposite signs.  
Hence, it is unclear whether the transition is real or just an artifact.  To illustrate, suppose that Riess \etal had found $z_t=3$; since their farthest SNeIa has $z=1.76$, it would have been clear that $z_t$ was implied not by the data but by the model itself.  Since $z_t$ actually does lie in the observed redshift range, evidence for a transition exists but remains unconvincing.

An alternate parametrization is a piecewise constant acceleration with two distinct epochs \cite{Turner:2001mx}:
\be
q(z)=\left\{
\begin{array}{ll}
  q_0 & \mbox{ for } z\leq z_t  \\
  q_1 & \mbox{ for } z>z_t,  \label{eq:q2epoch}
\end{array}
\right.
\ee
where $q_0$ and $q_1$ correspond to the average values of $q$ in their respective epochs.  \rf{q0q1} shows our confidence contours in the $q_0-q_1$ plane for various values of $z_t$.  The best fit model has $z_t=0.08$, $q_0=-2.0$, $q_1=0.11$, 
$\chisq/{\rm DOF}=173/153$ and a 13\% goodness-of-fit; however,
the $\chisq$ minimum in $z_t$ is broad and the constraints on $q_0$ and $q_1$ vary significantly with $z_t$.  Moreover, in the limit of $z_t \ll 0.1$ ($z_t \gg 1$), the model becomes essentially a one-zone model as there are no SNe to constrain $q_0$ ($q_1$) and the error on $q_0$ ($q_1$) blows up.  Marginalizing over $z_t$ with a prior that takes this into account, $z_t\in[0.1,1]$, one obtains the final constraint contour in \rf{q0q1}. Again there is strong evidence for recent acceleration (negative $q_0$ contains over 99.9\% of the probability) and weaker evidence for past deceleration (positive $q_1$ contains 91\% of the probability).  
The marginalized distribution for $z_t$ is very broad (see \rf{hist2epoch}) because $q_0$ and $q_1$ are poorly constrained for low and high $z_t$, and meaningful constraints to $z_t$ cannot be obtained.  

\begin{figure}[tp]
  \sfig{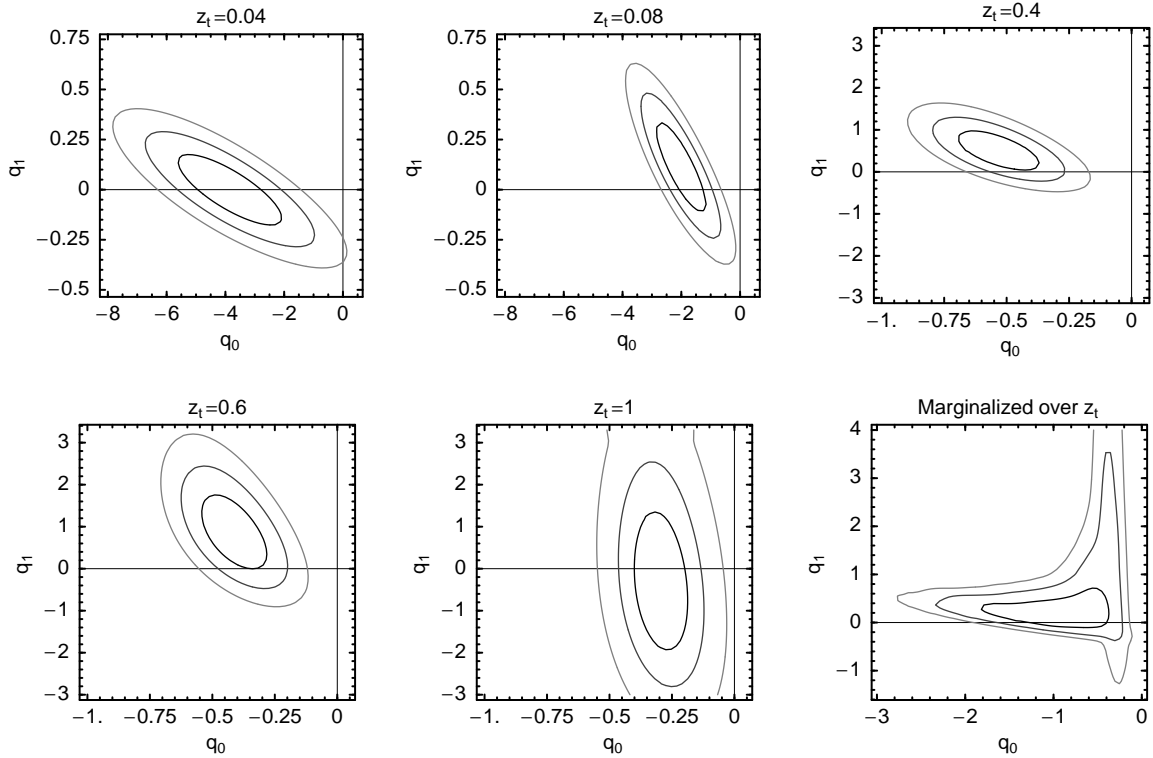}{.9\columnwidth}
  \caption{{\small Confidence contours in the $q_0-q_1$ plane for the
  two-epoch model.  The 68.3\%, 95.4\% and 99.7\% contours are shown for various
  values of $z_t$.  The $\chisq$ function is broad in the $z_t$ direction and minimized for $z_t=0.08$.  The final frame shows the contours after marginalization over $z_t\in[0.1,1]$. }}
  \label{fig:q0q1}
\end{figure}

As shown in \rf{allqplot}, the best-fitting kinematic models appear different from \LCDM -- in particular, they both find a lower redshift for the transition to acceleration.  
To address the significance of this, we generated 1000 mock SNeIa data sets, assuming \LCDM with $\Omega_M=0.31$ and each with 157 magnitudes and errors taken from the gold set.  We fit both kinematic models to the simulated data sets and tabulated the best-fit parameters.  \rf{hist2epoch} and \rf{histlinear} show that the kinematic parameters we expect to measure in a \LCDM Universe are consistent with the constraints given by the gold set.  {\it If} there is a difference between the actual acceleration history of the Universe and \lcdm, these kinematic models and the present data do not have the statistical power to demonstrate it.

\begin{figure}[thbp]
  \sfig{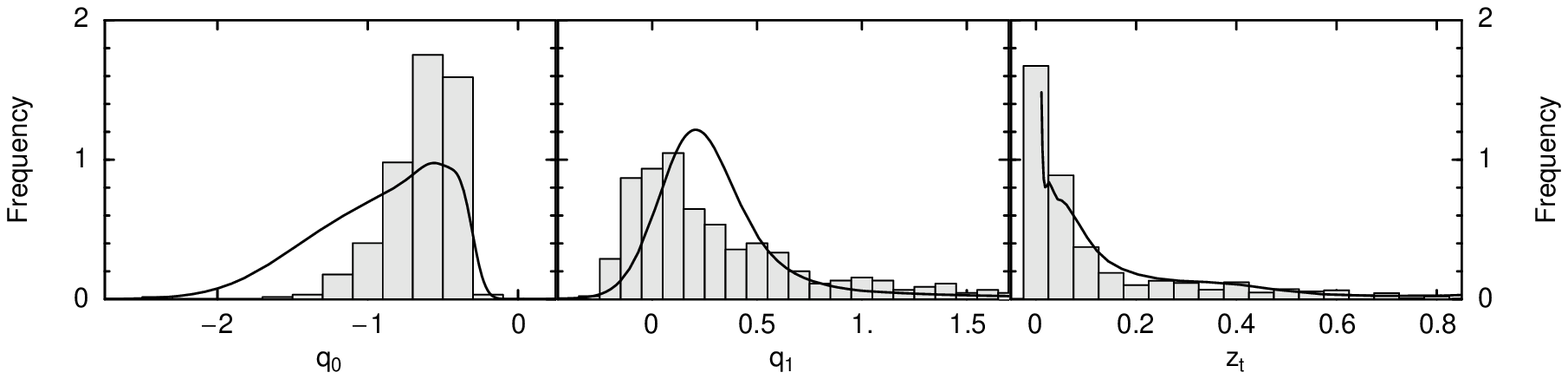}{.9 \columnwidth}
  \caption{{\small Constraints on the parameters of the 2-epoch model are compared to their expected measurements in a \LCDM Universe.  The solid lines are probability density functions obtained from the gold set and the histograms were generated from 1000 mock datasets as described. The $q_0$ ($q_1$) function and histogram were marginalized over $q_1$ ($q_0$) and $z_t\in[0.1,1].$  The $z_t$ function and histogram were summed over the full range of $q_0$ and $q_1$.}}
  \label{fig:hist2epoch}
\end{figure}

\begin{figure}[tp]
  \sfig{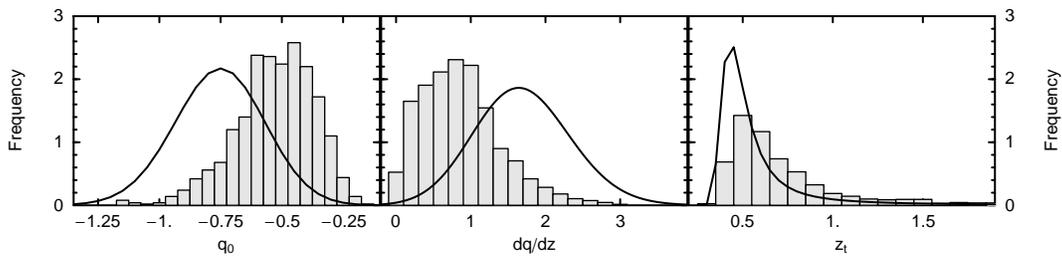}{.9 \columnwidth}
  \caption{{\small Constraints on the parameters of the linear model are compared to their expected measurements in a \LCDM Universe.  The solid lines are probability density functions obtained from the gold set and the histograms were generated from 1000 mock datasets as described. The transition redshift is not an independent parameter; it is given by $z_t=-q_0/(dq/dz)$.}}
  \label{fig:histlinear}
\end{figure}

\vspace{1cm}
Both the 2-epoch and linear $q(z)$ models provide direct evidence for a recent period of acceleration and weaker evidence for a past period of deceleration.  However, as discussed, they each have their shortcomings.  To further illustrate, we ask the provocative question, could the universe be decelerating today?  First, as a practical matter, the Hubble flow only begins to dominate local peculiar motions for $z > 0.01$; this implies that any knowledge of the expansion dates back to at least $0.01H_0^{-1} \sim 0.1\,$Gyr -- so as far as we know, the Universe could have stopped expanding 100\,Myr ago!  Next, consider the standard Taylor expansion of the luminosity distance,
\begin{equation}
d_L = H_0^{-1}z + {1\over 2}(1-q_0)H_0^{-1}z^2 + \ldots
\end{equation}
Because the acceleration term is second order in $z$, one is further limited in detecting its effect by the accuracy of measuring $d_L$:
 10\% uncertainty in $H_0\,d_L$ implies that objects at redshift greater than $z\sim 0.1$ are needed to measure $q_0$ with an error of order unity.  Thus, direct evidence of acceleration/deceleration more recent than 1\,Gyr ago will be difficult to obtain; this can be seen quantitatively in \rf{q0q1} where for $z_t < 0.1$ the error ellipse on $q_0$ blows up.  

\begin{figure}[t]
  \sfig{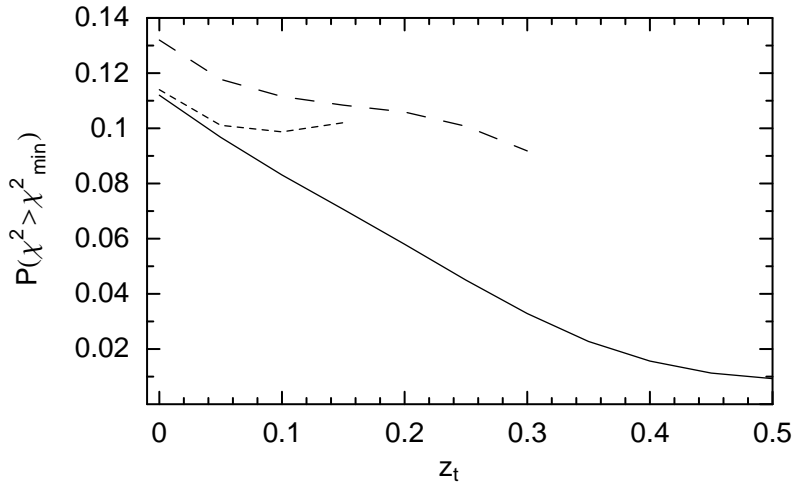}{.6\columnwidth}
  \caption{{\small Maximum goodnesses-of-fit for three-epoch $q(z)$ models that are not accelerating today.  Setting $q(z<z_t)=0$ and $q(z>z_e)=0.5$, the solid, long-dashed and short-dashed lines show $z_e=$ 0.5, 0.3 and 0.15 respectively. }}
  \label{fig:noaccel}
\end{figure}

The reason for asking whether or not the Universe is accelerating today is not just academic; there exist dark energy models that predict a history of alternating acceleration and deceleration \cite{Dodelson:2001fq,German:2004db}.  To address this issue we consider a three-epoch model,
\be
q(z)=\left\{
\begin{array}{ll}
  q_0 & \mbox{ for } z\leq z_t  \\
  q_1 & \mbox{ for } z_t<z<z_e  \\
  0.5 & \mbox{ for } z\geq z_e ,
\end{array} \right.
\ee
motivated by the evidence for a recent epoch of acceleration and an earlier epoch of deceleration. The parameters $q_0$ and $z_t$ allow us to address the question of whether or not the Universe has been accelerating recently.  To do so, we  
set $q_0=0$; \rf{noaccel} shows our goodness-of-fit as a function of $z_t$ after minimizing $\chisq$ with respect to $h$ and $q_1$.  With an early transition at $z_e=0.3$, a long epoch of recent deceleration is consistent with the data at the 10\% level.  Larger values of $q_0$ will decrease the goodness-of-fit, but note that there may exist better-fitting parametrizations that raise it.  In summary, the present SNeIa data cannot rule out the possibility that the Universe has actually been decelerating for the past 3 Gyr (i.e., since $z = 0.3$).

\subsection{Principal Component Analysis}\label{sec:pc}

Here, we follow Huterer and Starkman \cite{Huterer:2002hy} and let the data itself guide our parametrization.  By computing a Fisher matrix \cite{Dodelson:2003ft, Bond:1998zw}, we can find accurately measured parameters and use them to form robust conclusions about cosmic acceleration.  We start by dividing the redshift range of the SNeIa into $N$ bins of width $\Delta z$ and write
\be
q(z) = \sum_{i=1}^N \beta_i c_i(z)
\ee
where the $c_i(z)$ equal unity inside the $i$th bin and zero outside (NB: as $N\rightarrow \infty$, arbitrary $q(z)$ can be specified).  From the gold set and \re{mu}, it is straightforward to numerically calculate the  $(N+1)$x$(N+1)$ Fisher matrix $F$ for the $\beta_i$ and the Hubble parameter $h$.  For simplicity, we calculate $F$ using $\beta_i=0$ for all $i$; however, we have found that $F$ is rather insensitive to the choice of fiducial parameters.
After marginalizing $F$ over $h$, we find a new basis of orthonormal functions $e_i(z)$ which diagonalize $F$.  The deceleration rate can be expanded in terms of these modes or ``principal components'':
\be
q(z) = \sum_{i=1}^N \alpha_i e_i(z) \label{eq:qinpcbasis}.
\ee
The eigenvalues of $F^{-1}$ estimate the uncertainties $\sigma(\alpha_i)$ in the coefficients $\alpha_i$.  The advantage of the principle component basis is that
the errors on the $\alpha_i$ are uncorrelated.  With $\Delta z=0.05$, the 6 modes of $q(z)$ with the smallest errors are shown in \rf{qmodes}.
\begin{figure}[p]
  \sfig{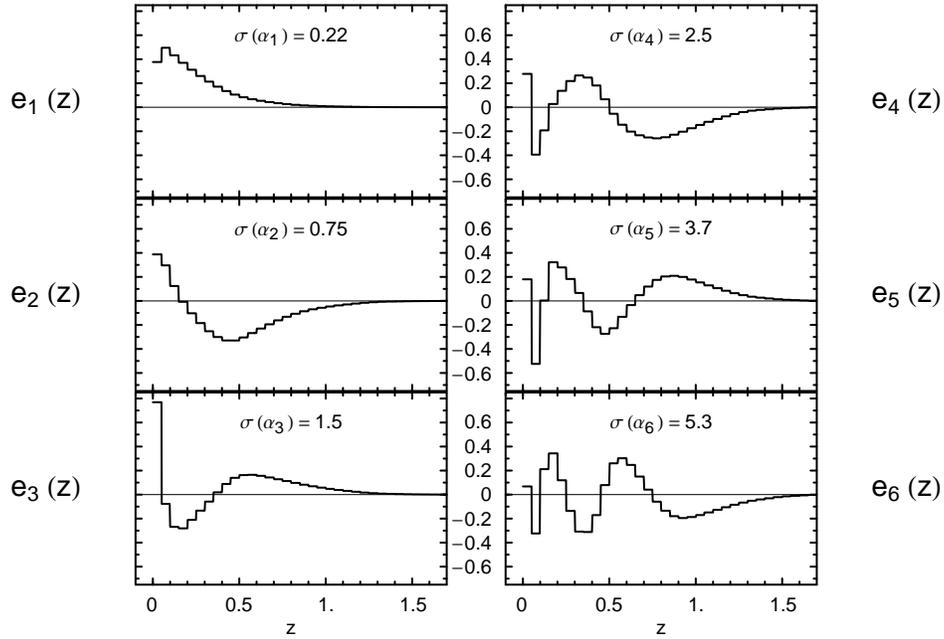}{.7\columnwidth}
  \caption{{\small The six most well-constrained principal components of $q(z)$.  They are normalized so that $e_i(0)>0$ and $\int e_i(z)^2\, dz$=1.  The errors on the mode amplitudes, $\sigma(\alpha_i)$, are estimated from the Fisher matrix.}}
  \label{fig:qmodes}
\end{figure}

We can use the principal components to reconstruct $q(z)$, i.e. using \re{qinpcbasis} as our model, we can fit the $\alpha_i$ to SNeIa data.  
Since $F$ is insensitive to fiducial parameters, we may keep the $\sigma(\alpha_i)$ as error estimates.  Note that if $F$ were sensitive to fiducial parameters, we would have to ensure that the best-fit model was the same as the fiducial model before calculating the $\sigma(\alpha_i)$; such a model can be found iteratively.
Table \ref{tab:recon} shows the results of parameter fits which use progressively more modes, starting with the most well-constained.  The 1st and 2nd mode coefficients are respectively about 5-$\sigma$ and 2-$\sigma$ negative  while higher modes are consistent with zero due to increasingly large variances.  Increasing the number of modes slightly alters the fit; however, since the data is insensitive to higher modes and since the coefficients are uncorrelated, we find it unlikely that keeping more modes will change the coefficients in Table \ref{tab:recon} by 1-$\sigma$.
\begin{table}[thbp]
\begin{center}
\begin{tabular}{|c|cccccc|}\hline
\# modes&$\chisq$&h&$\alpha_1$&$\alpha_2$&$\alpha_3$&$\alpha_4$\\\hline
1 & 177.7 & .644 & -1.10 & --    & --    & --   \\
2 & 173.2 & .653 & -1.08 & -1.61 & --    & --   \\
3 & 172.6 & .664 & -1.08 & -1.71 & -1.10 & --   \\
4 & 172.6 & .662 & -1.08 & -1.56 & -1.30 & .61 \\ \hline \hline
$\sigma(\alpha_n)$ & &        & 0.22  & 0.72  & 1.52  & 2.54 \\ \hline
\end{tabular}
\caption{Reconstruction of $q(z)$ by \re{qinpcbasis} using well-constrained
principal components.  The top part of the table shows the results of parameter fits as progressively more modes are added.  The estimated errors on the parameters follow directly from the Fisher matrix used to derive the modes.  We have marginalized over $h$.}\label{tab:recon}
\end{center}
\end{table}

\begin{figure}[p]
  \sfig{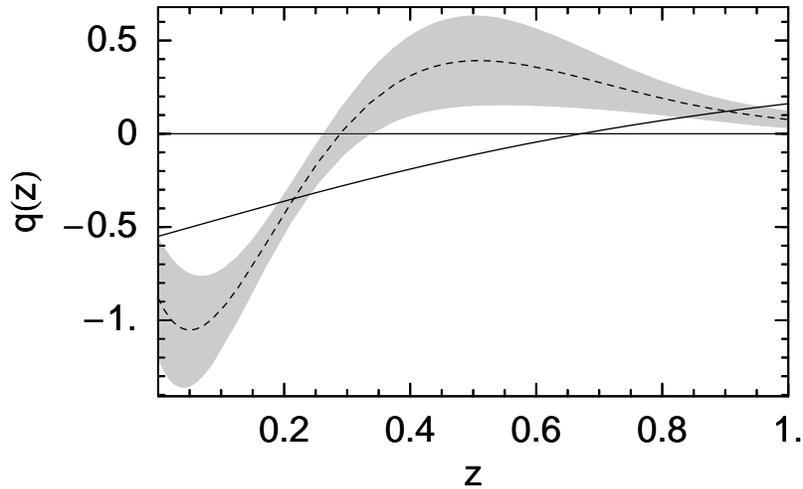}{.6\columnwidth}
  \caption{{\small Reconstruction of $q(z)$ using its 2 most well-constrained principle components (\re{qinpcbasis}).  The short-dashed line is the best 2 mode fit to SNeIa data; it has a 14\% goodness-of-fit.  The gray band shows the error corresponding to 1-$\sigma$ uncertainties in both $\alpha_1$ and $\alpha_2$.  The solid line is $q(z)$ for \LCDM with $\Omega_M=0.3$.  As in \rf{allqplot}, the kinematic model suggests a later transition to acceleration than \LCDM does.  }}
  \label{fig:2modefit}
\end{figure}

It is sensible to reconstruct $q(z)$ using only the well-measured modes \cite{Huterer:2002hy}.  Doing so gives us a coarse yet accurate picture of the deceleration parameter in its ``mode space,'' but it also introduces a bias: the reconstruction will lack features of the true $q(z)$ which are poorly probed by SNeIa.  Consider for instance the two-mode reconstruction illustrated in \rf{2modefit}; it approaches zero at high redshift, but it is clear that we cannot trust this feature.  It is also clear that the reconstruction will look different if we include more modes.  
So which features of the reconstruction \emph{can} we trust?


First, we can trust that the Universe has accelerated.  To see this, we note that the 1st mode is completely positive.  Since the modes are orthonormal, the coefficients $\alpha_i$ are given by
\be
\alpha_i = \int dz\; q(z)e_i(z) \label{eq:coef} .
\ee
It follows that $\alpha_1$ is negative only if $q(z)<0$ for some $z$.  Our estimate that $\alpha_1$ is 5-$\sigma$ negative therefore implies a 5-$\sigma$ detection of cosmic acceleration \emph{at some redshift}.  The shape of $e_i(z)$ implies that this redshift is likely to be low.
This conclusion is not model-dependent since we are free to decompose $q(z)$ using any complete basis we like.  One could make a similar argument using e.g. a polynomial or Fourier expansion, but the principle component expansion is advantaged since the uncertainty in $\alpha_1$ is small and non-degenerate with other parameters.

\begin{figure}[tp]
  \sfig{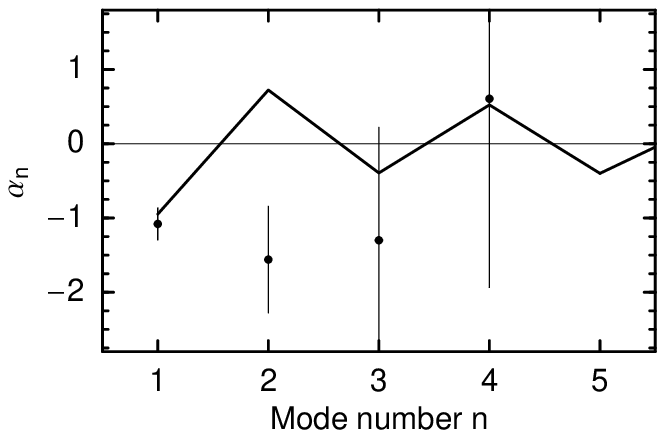}{.5\columnwidth}
  \caption{{\small Mode space representation of $q(z)$.  The solid line indicates the $\alpha_n$ corresponding to $q(z)=-0.29$.  The points show the values of $\alpha_n$ obtained from a 4-mode fit, and the error bars are taken from the Fisher matrix (see Table \ref{tab:recon}).}}
  \label{fig:modespace}
\end{figure}

Next, we can also trust that $q(z)$ is not constant.  Fitting a constant $q$ model to the gold set yields $q=-0.29$ with a 6\% goodness-of-fit, but fitting only to modes that have been convincingly measured shows that constant $q$ really fails.  Using the reconstructions in Table \ref{tab:recon}, we fit a constant $q$ model directly to the mode coefficients.  The best fit to these reconstructions is also $q=-0.29$; it is essentially determined by the first two modes, which have the smallest errors.  We obtain the following goodnesses-of-fit:
\be \nonumber
P(\chisq>\chisq_{\rm min}) =
\left\{
\begin{array}{ll}
0.1\% & \mbox{ first 2 modes} \\
0.4\% & \mbox{ first 3 modes} \\
1\% & \mbox{ first 4 modes}
\end{array}
\right.
\ee
The main reason for the poor fit is easily seen in \rf{modespace}:  if $q$ were constant, the first two coefficients should have roughly opposite values, but they don't.  Higher modes will tend to improve the fit by adding degrees of freedom without significantly increasing $\chisq$.

\section{Conclusions}
Understanding dark energy may ultimately require us to abandon or change the familiar equations of general relativity.  Hence it is crucial that we keep track of which observations and conclusions depend on the assumption of general relativity and which are more robust.  Without the Friedmann equation, it is still possible to describe the Universe with a Robertson-Walker metric and infer the expansion history via the dimming and cosmological redshifting of SNeIa.  We used this fact to learn about cosmic acceleration kinematically.

In particular, we have shown that the gold set of SNeIa from Riess \etal provides the following information:

\begin{itemize}
\item {\bf Very strong evidence that the Universe once accelerated --}  Though various parametrizations can imply this, the strongest evidence comes from measuring the best-constrained principal component of $q(z)$.  The acceleration is likely to have been relatively recent in cosmic history.
\item {\bf Strong evidence that $q$ was higher in the past --}  The constant $q$ model has a poor goodness-of-fit, especially when fit in mode space using only well-measured modes.  In addition, the linear $q(z)$ model indicates that the average $dq/dz$ is positive.
\item {\bf Weak evidence that the Universe once decelerated --}  Both the two-epoch and two-mode models of $q(z)$ slightly favor past deceleration.  The linear model also predicts a transition to positive $q$.  Alas, it is difficult to show that this is not a model-dependent feature.
\item {\bf Little or no evidence that the Universe is presently accelerating --}  It is manifestly difficult to constrain $q(z<0.1)$ with SNeIa.  Using a three-epoch model,
we further showed that a long period (e.g., $z= 0 \rightarrow 0.3$) of recent deceleration is consistent with data.  
\item {\bf \LCDM and $w$CDM are acceptable fits to SNeIa data.}  
While there are tantilizing indications that the actual acceleration history may be different, none of the kinematic models studied here have revealed robust features about cosmic acceleration that differ from \LCDM or $w$CDM.  Moreover, none of the kinematic models fit the data significantly better.  
\end{itemize}

\begin{acknowledgments} 
We thank Dragan Huterer and Eduardo Rozo for many valuable discussions.  Thanks also to Paul Abbott for an insightful programming suggestion.  CAS was supported in part by the Department of Energy, the Kavli Institute for Cosmological
Physics at the University of Chicago and by NSF grant PHY-011442.
\end{acknowledgments}

\bibliography{sneq_dec22}

\end{document}